\documentclass[aps,prl,reprint,twocolumn,superscriptaddress]{revtex4-1}
\usepackage{amsmath}
\usepackage{amsfonts}
\usepackage{amssymb}
\usepackage{amsthm}
\usepackage{graphicx}
\usepackage{CJK}
\usepackage{times}
\usepackage{mathptmx}
\usepackage[colorlinks, linkcolor=blue, urlcolor=blue, anchorcolor=blue, citecolor=blue]{hyperref}
\usepackage{mathtools}
\usepackage{lipsum}
\usepackage{array}
\usepackage[table]{xcolor}
\definecolor{mygray}{gray}{.9}

\begin{document}


\title{Spin-wave Goos-H\"{a}nchen effect induced by 360 degree domain walls in magnetic heterostructures}



\author{Mei Li}
\affiliation{College of Physics Science and Technology, Yangzhou University, Yangzhou 225002, People's Republic of China}
\author{Bin Xi}
\affiliation{College of Physics Science and Technology, Yangzhou University, Yangzhou 225002, People's Republic of China}
\author{Yongjun Liu}
\affiliation{College of Physics Science and Technology, Yangzhou University, Yangzhou 225002, People's Republic of China}
\author{Jie Lu}
\email{lujie@yzu.edu.cn}
\affiliation{College of Physics Science and Technology, Yangzhou University, Yangzhou 225002, People's Republic of China}


\date{\today}

\begin{abstract}
In this work, lateral displacements of transmitted and reflected spin waves at a 360 degree domain wall (360DW), 
which is referred to as the spin-wave Goos-H\"{a}nchen effect (SWGHE),
are systematically investigated in magnetic heterostructures with perpendicular easy/hard axis and wall-extension direction.
Similar to the counterpart at heterochiral interfaces, the interfacial Dzyaloshinskii-Moriya interactions (IDMI)
originating from a heavy-metal substrate is important for the emergence of SWGHE.
More interestingly, the SWGHE can even survive in ferromagnets with biaxial anisotropy 
(either intrinsic or caused by shape anisotropy) in the absence of IDMI
due to the unique 360DW-induced potentials which are distinct to the well-known P\"{o}schl-Teller ones.
Numerics shows that these lateral displacements are generally fractions of the spin-wave wavelength.
They can be further enhanced by an array of well-separated 360DWs thus provide a large variety 
for spin-wave manipulation.
\end{abstract}


\maketitle


\section{I. Introduction} 
Magnonics, devoting to the behaviors and applications of magnons in magnetic materials, 
has developed rapidly 
in recent years due to the advantages of magnons such as the nanoscale wavelength, 
the sub-terahertz frequency and the absence of Joule heating\cite{Grundler_JPDAP_2010,Hillebrands_JPDAP_2010,Munzenberg_PhysRep_2011,Hillebrands_NatPhys_2015,Schultheiss_PhysRep_2021,YanPeng_PhysRep_2021,YanPeng_PhysRep_2023}. 
Manipulations on the various degree of freedom of magnons, such as the amplitude,
wave vector, phase, and polarization, lead to the development of extremely rich 
magnetic nanodevices, for example, the spin-wave amplifiers\cite{Ono_PRL_2009,Rezende_PRL_2011,XQLi_PRB_2014,Beach_JAP_2017,Demokritov_AdvMater_2018,Pertsev_PRB_2021,Demidov_nc_2024,XRWang_APL_2024},
lens\cite{Hansen_srep_2016,Csaba_IEEE_2018,YanPeng_PRB_2020,Demokritov_APL_2016,Kruglyak_APL_2018,Freymann_APL_2020,CLiu_JMMM_2022,Kruglyak_PRB_2015,Freymann_srep_2018,Krawczyk_Nanoscale_2019,Schutz_PRB_2020}, 
phase shifters\cite{Kruglyak_APL_2012,ZZhong_JMMM_2015,Slavin_AIPAdv_2016,Berakdar_JAP_2018,ZZhong_JPDAP_2020,JZhao_AIPAdv_2021},
and polarizers/analyzers\cite{JLan_nc_2017,JLan_PRB_2021,Vignale_PRB_2022}, etc. 

The Goos-H\"{a}nchen effect\cite{GH_original_paper_1947} is a basic phenomenon in optics, saying that 
when a beam of light is incident on the interface between two media, the 
reflected and transmitted lights respectively suffer a lateral displacement.
Further studies show that the Goos-H\"{a}nchen effect is a manifestation
of the wave nature of light, thus should have counterparts in other waves,
such as acoustic\cite{Lamkanfi_APL_2008}, electronic\cite{CFLi_JOpt_2013}, and neutron\cite{Langridge_PRL_2010} waves. 
For spin waves, existing investigations have confirmed the existence of 
Goos-H\"{a}nchen effect at the interface between two ferromagnetic (FM) films\cite{Krawczyk_APL_2012,Krawczyk_PRB_2017,Krawczyk_IEEE_2017,YanPeng_PRB_2019,Krawczyk_JMMM_2019,DDeng_Opt Commun_2020,XFHan_PRB_2021}
or the edge of a single ferromagnet\cite{Krawczyk_APL_2014,Krawczyk_PRB_2015,Ono_PRL_2018}. 
In addition, due to the abundance of magnetic texture in ferromagnets
the spin-wave Goos-H\"{a}nchen effect (SWGHE) can even emerge at 
one-dimensional (1D) magnetic solitions\cite{Winter_PR_1961,Goll_JPCS_2010,Stamps_AEM_2016,Laliena_srep_2020,Laliena_AEM_2022,Laliena_PRB_2022}.
The pioneer theoretical work on SWGHE at 1D solitons is perturbative\cite{Stamps_AEM_2016}.
From 2020, in a series of works Laliena $\mathit{et. al.}$ have developed a 
non-perturbation theory and systematically studied the SWGHE 
at a Sine-Gordon soliton in an ideal 1D monoaxial helimagnet\cite{Laliena_srep_2020,Laliena_AEM_2022,Laliena_PRB_2022}.
However, to our knowledge the corresponding study on SWGHE 
at a 360 degree domain wall (360DW) in FM layers 
of real magnetic heterostructures (MHs), especially with perpendicular easy/hard axis and wall-extension direction, is absent.

In the present work, with the help of non-perturbation theory 
we perform a thorough investigation on SWGHE in real MHs with 
heavy-metal substrates providing the interfacial Dzyaloshinskii-Moriya 
interactions (IDMI)\cite{Dzyaloshinsky,Moriya,Bogdanov_JMMM_1994} which is important for the emergence of SWGHE.
The lateral displacements are first theoretically deduced and then numerically 
confirmed to be sub-wavelength of the spin waves involved.
Different from the existing assertion, the SWGHE at 360DWs can even survive 
in ferromagnets with biaxial anisotropy (either intrinsic or caused by shape anisotropy) in the absence of IDMI
owing to the unique 360DW-induced potentials which is distinct to the 
most-commonly-encountered P\"{o}schl-Teller ones\cite{Poschl_Teller_1933}. 

The rest of this paper is organized as follows. 
In Sec. II, the system set up and its modelization are briefly introduced. 
In Sec. III, by ignoring the nonlocal component of magnetostatic energy 
a rigorous static 360DW profile is rediscovered and turns out to be identical with
that from an effective analysis for FM films\cite{Slastikov_PRSA_2005,Muratov_JCP_2006,Muratov_JAP_2008}.
Then its static and dynamical stability are explored in Secs. III and IV, respectively.
In particular, the spin wave operators (essentially from the perturbation of static 360DW profile), 
as well as the corresponding (semi)positive-definite regions are obtained, which constitute the 
basis of non-perturbation theory for spin waves.
With these results in hand, the spin-wave spectral problem under 360DW potential are solved in Sec. V.
Based on all above preparations, the SWGHE occurs at the 360DW is systematically 
presented in Sec. VI within the scattering theory framework.
At last, discussions and conclusion are drawn in Sec. VII.

\section{II. Modelization} 

\hspace*{\fill} 
~\\
~\\
\begin{figure} [htbp]
	\centering
	\includegraphics[width=0.40\textwidth]{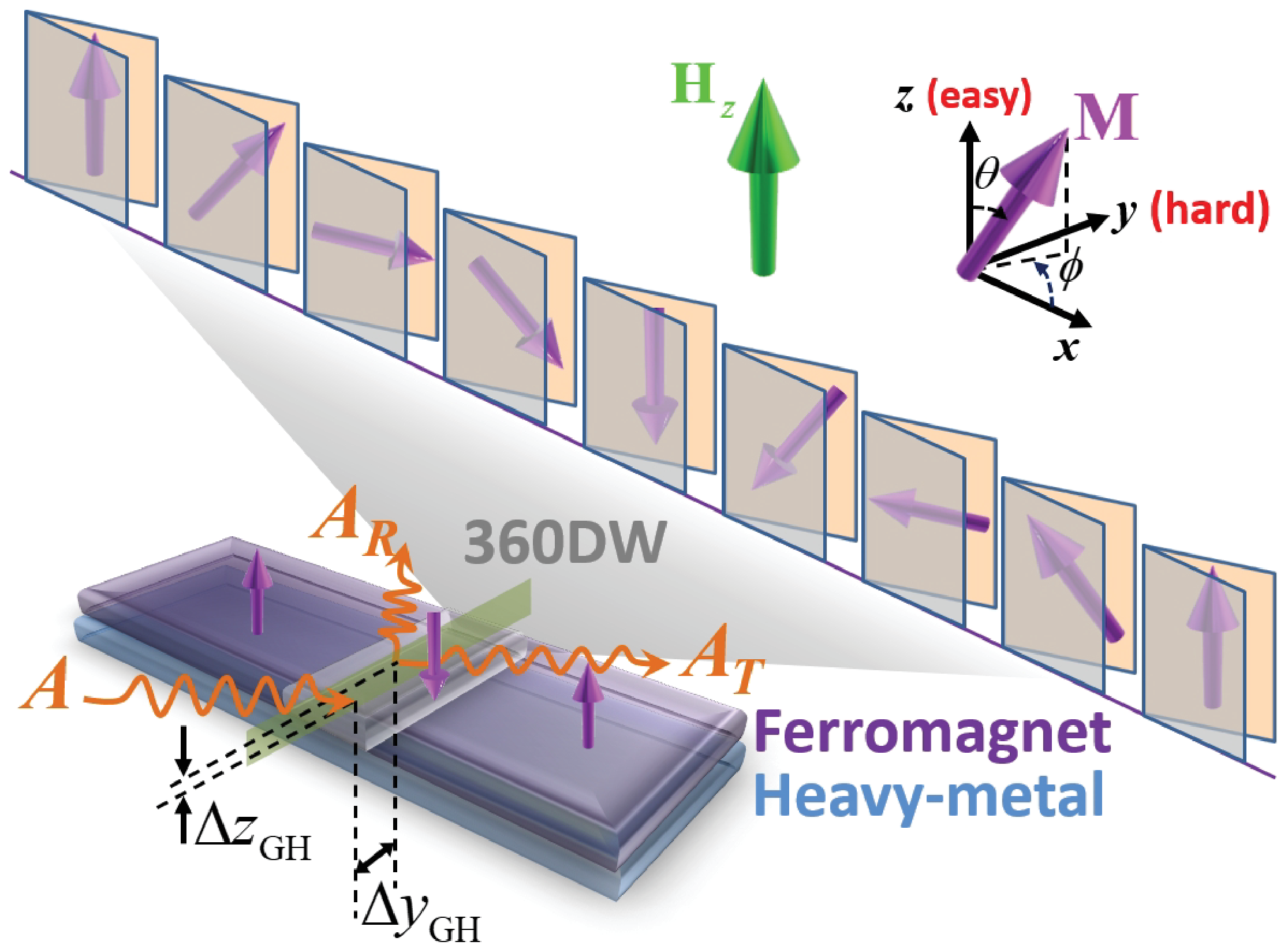}  %
	\caption{(Color online)  Illustration of the spin-wave Goos-H\"{a}nchen effect (SWGHE) 
		at a 360DW (gray region) in the (purple) FM layer of a MH with the (blue) heavy-metal substrate.
		The global Cartesian coordinate system is indicated at the rightup corner and
		the magnetization $\mathbf{M}=M_s \mathbf{m}$ is fully described by its polar
		and azimuthal angles ($\theta$ and $\phi$).
		An external magnetic field is applied along the easy axis of the FM layer
		to assure the existence and further manipulate the wall profile.
		An incident spin wave with amplitude $A$ is scattered at the 360DW.
		The outgoing reflected (transmitted) wave bearing an amplitude $A_{R}$ ($A_T$) 
		suffers lateral displacements $\Delta y_{\mathrm{GH}}$ and $\Delta z_{\mathrm{GH}}$.
	}\label{fig1}
\end{figure}

We consider a MH composed of a heavy-metal substrate supplying IDMI
and a FM layer with perpendicular magnetic anisotropy (PMA) 
capable of hosting various magnetic solitons and spin waves, 
as shown in Fig. \ref{fig1}.
The easy axis lies in $z-$axis, the long axis of the MH is taken as $x-$axis along which 1D magnetic solitons 
extend, correspondingly $\mathbf{e}_y=\mathbf{e}_z\times\mathbf{e}_x$.
The magnetic energy density $\mathcal{E}[\mathbf{m}]$ of the FM layer in this MH
is a functional of the magnetization unit vector field $\mathbf{m}(\mathbf{r})$ and
includes four parts: 
the exchange term $\mathcal{E}_{\mathrm{ex}}=A(\mathbf{\nabla}\mathbf{m})^2$ 
($A$ being the exchange stiffness), 
the Zeeman term $\mathcal{E}_{\mathrm{Zee}}=-\mu_0 M_s\mathbf{m}\cdot\mathbf{H}_a$ with
the external field $\mathbf{H}_a=H\mathbf{e}_z$ and the saturation magnetization $M_s$,
the anisotropy term $\mathcal{E}_{\mathrm{ani}}=(\mu_0 M_s^2/2)(-k_{\mathrm{E}}m_z^2+k_{\mathrm{H}}m_y^2)$ where $k_{\mathrm{E}}$ ($k_{\mathrm{H}}$) is the total (crystalline plus shape) anisotropy coefficient in easy (hard) axis, 
and the IDMI contribution\cite{Bogdanov_JMMM_1994}  $\mathcal{E}_{\mathrm{IDMI}}=D_{\mathrm{i}}[m_z\mathbf{\nabla}\cdot\mathbf{m}-(\mathbf{m}\cdot\mathbf{\nabla})m_z]$
with $D_{\mathrm{i}}$ being the IDMI strength.
In the analytical part of this work we ignore the nonlocal component of the magnetostatic energy,
which has been proved to be effective in existing analytical works.

The dynamics of $\mathbf{m}(\mathbf{r})$ is described by the
Landau-Lifshitz-Gilbert (LLG) equation\cite{LLG_equation}
\begin{equation}\label{LLG_equation}
	\partial_t \mathbf{m}=-\gamma \mathbf{m}\times \mathbf{H}_{\mathrm{eff}} + \alpha\mathbf{m}\times\partial_t \mathbf{m},
\end{equation}
in which $\gamma$ and $\alpha$ are respectively
the electron gyromagnetic constant and the Gilbert damping coefficient.
The effective field,
$\mathbf{H}_{\mathrm{eff}}=-(\mu_0 M_s)^{-1}\delta\mathcal{E}/\delta\mathbf{m}$, turns to be
\begin{equation}\label{H_eff_general}
	\mathbf{H}_{\mathrm{eff}}=\frac{2A}{\mu_0 M_s}\nabla^2\mathbf{m}+M_s\left(k_{\mathrm{E}}m_z-k_{\mathrm{H}}m_y\right)+\mathbf{H}_a +\mathbf{H}_{\mathrm{IDMI}},
\end{equation}
with
\begin{equation}\label{H_IDMI_general}
	\mathbf{H}_{\mathrm{IDMI}}=-\frac{2D_{\mathrm{i}}}{\mu_0 M_s}\left[\left(\nabla\cdot\mathbf{m}\right)\mathbf{e}_z-\nabla m_z\right].
\end{equation}

\section{III. Static stability of 360DWs} 

We first explore the rigorous profile of possible 1D 360DWs extending along the $x-$axis of this MH.
The unit vector along the direction that the polar (azimuthal) angle of $\mathbf{m}$ increases
is denoted as $\mathbf{e}_{\theta}$ ($\mathbf{e}_{\phi}$).
Then $(\mathbf{m},\mathbf{e}_{\theta},\mathbf{e}_{\phi})$ forms a local orthonormal triad.
A static magnetic soliton requires that $\mathbf{m}\parallel\mathbf{H}_{\mathrm{eff}}$,
which is equivalent to $\mathbf{H}_{\mathrm{eff}}\cdot\mathbf{e}_{\theta}=\mathbf{H}_{\mathrm{eff}}\cdot\mathbf{e}_{\phi}=0$, holds everywhere.
In the assumptions of 1D system and no twisting in the azimuthal angle ($\partial_x \phi=0$),
$\mathbf{H}_{\mathrm{eff}}\cdot\mathbf{e}_{\phi}=0$ leads to:
$\sin\theta\sin\phi[2D_{\mathrm{i}}(\mu_0 M_s)^{-1}\partial_x\theta-k_{\mathrm{H}}M_s\cos\phi]=0$.
For a 1D soliton, $\theta$ must be a nontrivial function of $x$. 
Thus one has $\sin\phi=0$, i.e. $\phi=n\pi$ with $n$ being any integer.
Whether $n$ is even or odd depends on the sign of IDMI strength, implying 
that IDMI brings definite chirality to the static 360DW, as will be illustrated later.
Furthermore, $\mathbf{H}_{\mathrm{eff}}\cdot\mathbf{e}_{\theta}=0$ gives
\begin{equation}\label{Equation_for_360DW}
	\frac{d^2\theta}{dX^2}-\frac{1}{2}\sin 2\theta-h\sin\theta=0,
\end{equation}
with $h\equiv H/(k_{\mathrm{E}}M_s)$, $X\equiv x/\Delta_0$ and 
$\Delta_0\equiv\sqrt{2A/(\mu_0 k_{\mathrm{E}}M_s^2)}$.

Equation (\ref{Equation_for_360DW}) can be solved analytically. 
Before that, we would like to point out that the polar and azimuthal angles 
in the usual spherical coordinates are not mathematically convenient 
to adequately describe a 360DW 
(for example, a ``$\uparrow\rightarrow\downarrow\leftarrow\uparrow$" type wall in the PMA system)
due to the reciprocation in polar angle and a jump of $\pi$ in azimuthal angle 
when the unit magnetization vector passes the south pole, hence
artificially bringing a discontinuity in the exchange energy density.
To fix this, a monotonic ``0 to $2\pi$" $\vartheta$ and a fixed $\varphi$ are adopted\cite{jlu_JMMM_2021}.
When the unit magnetization vector exceeds the south pole and goes back to the north pole,
$2\pi-\vartheta$ and $\varphi+\pi$ are respectively the polar and azimuthal angles in the 
usual spherical coordinate system.
This can be confirmed by the same magnetization components $m_{x,y,z}$ provided by
the two sets ``$(\vartheta,\varphi)$" and ``$(2\pi-\vartheta,\varphi+\pi)$".
Based on the above preparations, the rigorous solution to Eq. (\ref{Equation_for_360DW}) is
\begin{equation}\label{Profile_of_360DW}
	X=\frac{-1}{\sqrt{1+h}}\sinh^{-1}\left(\sqrt{\frac{1+h}{h}}\cot\frac{\vartheta}{2}\right),
\end{equation}
with $\sinh^{-1}(\rho)\equiv\ln(\rho+\sqrt{\rho^2+1})$.

We would like to make some remarks about the 360DW profile in Eq. (\ref{Profile_of_360DW}).
First, it is totally different from the famous Sine-Gordon soliton $X=h^{-1/2}\ln[\tan(\vartheta/4)]$
which is the solution of the equation: $d^2\theta/dX^2-h\sin\theta=0$\cite{Laliena_AEM_2022}. 
Obviously, the term ``$(1/2)\sin 2\theta$" in Eq. (\ref{Equation_for_360DW}) 
originating from the ``easy $z$-axis" anisotropy (perpendicular to $x-$axis that 360DWs extend) 
results in this difference.
Second, the influence of IDMI does not appear explicitly in Eq. (\ref{Profile_of_360DW})
due to the cancellation of IDMI terms in expanding $\mathbf{H}_{\mathrm{eff}}\cdot\mathbf{e}_{\theta}=0$.
However, it plays a decisive role in determining the chirality of 360DWs.
To see this, we examine the total magnetic energy density on the basis of  $\sin\varphi=0$
and find that only the IDMI term, that is $D_{\mathrm{i}}\partial_x\theta\cos\varphi$,
explicitly depends on $\varphi$.
The total IDMI energy turns to be $2\pi D_{\mathrm{i}}S\cos\varphi$
after integrating over the whole FM layer with $S$ being the cross section of the layer in $yz-$plane.
Obviously, it reaches its minimum when 
\begin{equation}\label{Chirality_of_360DW}
	\cos\varphi=-\mathrm{sgn(D_{\mathrm{i}})}\equiv \chi,
\end{equation}
with ``sgn" being the sign function, thus gives the static 360DW definite chirality.

In the following, we search for the possibility of realizing a ``chiral soliton lattice (CSL)" by a series of 
360DWs in this MH.  
After defining a dimensionless IDMI-strength
parameter $d_{\mathrm{i}}\equiv 2|D_{\mathrm{i}}|/(\mu_0 k_{\mathrm{E}}M_s^2 \Delta_0)$,
the difference in magnetic energy density of the static-360DW state and FM state,
$\Delta\mathcal{E}\equiv\mathcal{E}-\mathcal{E}_{\mathrm{FM}}$, turns to be
\begin{equation}\label{Energy_density_diff_between_360DW_FM}
	\frac{\Delta\mathcal{E}}{\mu_0 k_{\mathrm{E}}M_s^2}=\frac{\Delta_0^2}{2}\left(\partial_x\vartheta\right)^2+\frac{\sin^2\vartheta}{2}+h\left(1-\cos\vartheta\right)-\frac{d_{\mathrm{i}}\Delta_0}{2}\partial_x\vartheta.
\end{equation}
Then the magnetic energy difference, $\Delta E=\int d^3\mathbf{r}\Delta\mathcal{E}$, reads
\begin{equation}\label{Energy_diff_between_360DW_FM}
	\frac{\Delta E}{\mu_0 k_{\mathrm{E}}M_s^2\Delta_0 S}=f(h)-\pi d_{\mathrm{i}},
\end{equation}
with $f(h)\equiv4\sqrt{1+h}+2h\ln[(\sqrt{1+h}+1)/(\sqrt{1+h}-1)]$.

\begin{figure} [htbp]
	\centering
	\includegraphics[width=0.43\textwidth]{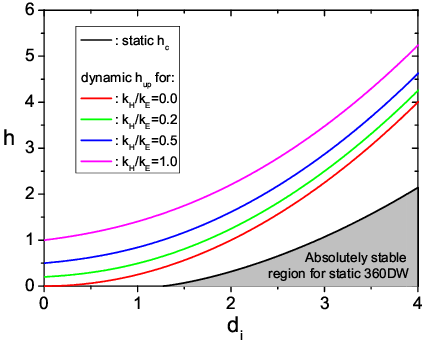}
	\caption{(Color online) Static and dynamical stability regions of 360DW in Eq. (\ref{Profile_of_360DW})
		in $(d_{\mathrm{i}},h)$ phase space.
		The black solid curve is the $h_c$ obtained from the critical condition $f(h)-\pi d_{\mathrm{i}}=0$.
		Correspondingly, the shaded (unshaded) area with $h<h_c$ ($h>h_c$) is the absolutely stable
		(metastable) region for the 360DW.
		As for the dynamical stability (under small-amplitude perturbation), the solid red, green, blue and
		magenta curves are $h_{\mathrm{up}}$ in Eq. (\ref{hc_definition}) with $k_{\mathrm{H}}/k_{\mathrm{E}}$
		equaling to 0.0, 0.2, 0.5, and 1.0, respectively. 
		The region with $h<h_{\mathrm{up}}$ is our best guess for the dynamically stable region 
		for the 360DW in Eq. (\ref{Profile_of_360DW}).}\label{fig2}
\end{figure}

Equation (\ref{Energy_diff_between_360DW_FM}) implies several interesting facts.
First, without IDMI ($d_{\mathrm{i}}=0$) the 360DW is always metastable since $f(h)>0$ always holds for positive $h$.
Second, the emergence of IDMI gives a certain chirality to the static 360DW thus 
ensures the minus sign in the right hand of Eq. (\ref{Energy_diff_between_360DW_FM}).
Then for a fixed strength $h$ of external field, under strong enough IDMI 
[$d_{\mathrm{i}}>f(h)/\pi$] the 360DW state becomes more stable than the FM state.
Therefore a CSL can be realized by arranging a series of 360DWs at sufficiently distant 
intervals along the long axis of MHs.
On the other hand, simple calculus tells us that the function $f(h)$ monotonically increases 
with $h$ and always larger than $f(h\rightarrow 0^+)=4$. 
Hence even in the presence of IDMI the 360DW state can also be metastable when 
$d_{\mathrm{i}}<4/\pi$.
In Fig. \ref{fig2}, the critical field strength $h_c$ from $\pi d_{\mathrm{i}}=f(h_c)$ that
separates the ``absolutely stable ($h<h_c$)" and ``metastable ($h>h_c$)" 
regions as IDMI increases is provided by the black solid curve. 

\section{IV. Dynamical stability of 360DWs} 

Except for the above analysis on static stability, we begin to analyze the dynamical stability
of the 360DW as well as the spin wave spectrum since the latter comes from
small-amplitude perturbation on the static 360DW.
We first denote $\mathbf{m}_0(x)$ as the wall profile shown in Eq. (\ref{Profile_of_360DW}).
With a small deviation from $\mathbf{m}_0$, the unit magnetization vector field
$\mathbf{m}$ can be written as
\begin{equation}\label{Deviation_of_m_from_m0}
	\mathbf{m}=\sqrt{1-\xi_1^2-\xi_2^2}\mathbf{m}_0 + \xi_1\mathbf{e}_1 + \xi_2\mathbf{e}_2,
\end{equation}
where $\mathbf{e}_1=\mathbf{e}_{\theta}$, $\mathbf{e}_2=\chi\mathbf{e}_y$ 
and $|\xi_{1,2}|\ll 1$.  
In the magnetization space we introduce a two-component vector
$\xi\equiv(\xi_1,\xi_2)^{\mathrm{T}}$, meantime in the real space we define the scalar product
\begin{equation}\label{Scalar_product_2x1}
	\left\langle\xi,\eta\right\rangle=\left(\xi_1,\eta_1\right)+\left(\xi_2,\eta_2\right),
\end{equation}
where $\left(f,g\right)=\int d^3\mathbf{R}f^{*}(\mathbf{R})g(\mathbf{R})$, 
and the dimensionless element of the reduced three-dimensional space reads
\begin{equation}\label{Full_dimensionless_space}
	d^3\mathbf{R}\equiv\frac{d^3\mathbf{r}}{\Delta_0^3}.
\end{equation}
Correspondingly, the total energy $E[\mathbf{m}]$ is expanded in powers of $\xi$ as
\begin{equation}\label{Energy_expansion}
	\frac{E[\mathbf{m}]-E[\mathbf{m}_0]}{\left(\mu_0 k_{\mathrm{E}}M_s^2/2\right)\Delta_0^3}= \left\langle\xi,\mathbb{K}\xi\right\rangle + O(\xi^3),
\end{equation}
in which the Hermitian operator $\mathbb{K}$ takes the form 
\begin{equation}\label{K_matrix}
	\mathbb{K}=
	\begin{bmatrix}
		\mathbb{K}_{11} & \mathbb{K}_{12} \\
		\mathbb{K}_{12}^{\dagger} & \mathbb{K}_{22}
	\end{bmatrix},
\end{equation}
where $\mathbb{K}_{12}=0$, $\mathbb{K}_{ii}=-\nabla_{\mathbf{R}}^2+U_i$, and
\begin{equation}\label{U1_U2_definitions}
  \begin{split}
  	U_1&=\cos 2\vartheta + h\cos\vartheta, \\
  	U_2&=d_{\mathrm{i}}\frac{d\vartheta}{dX}-\left(\frac{d\vartheta}{dX}\right)^2+\cos^2\vartheta+h\cos\vartheta+\frac{k_{\mathrm{H}}}{k_{\mathrm{E}}}.
  \end{split}
\end{equation}

A real large enough [compared with the magnetic exchange length
$\lambda\equiv\sqrt{A/(\mu_0 M_s^2)}$] magnetic material 
can be viewed as an infinite system, thus avoid the complications from varying boundary conditions.
Due to the fact that $E[\mathbf{m}]$, thus 
$\left\langle\xi,\mathbb{K}\xi\right\rangle$, is finite, the boundary condition
can be simply taken as: $\xi(\mathbf{r})$ vanishes rapidly enough 
as $|\mathbf{r}|\rightarrow+\infty$.
In the presence of any spin wave, the dynamical stability of the 360DW 
in Eq. (\ref{Profile_of_360DW}) requires that the Hermitian operator $\mathbb{K}$
should be positive semidefinite.
In the special case of $\mathbb{K}_{12}=0$ which is just the case in our work, this is equivalent to 
require that $\mathbb{K}_{11}$ and $\mathbb{K}_{22}$ are both positive semidefinite, and 
at least one of them is positive definite.

Interestingly, the Schr\"{o}dinger operator $\mathbb{K}_{11}$ with the potential $U_1$ presented in
Eq. (\ref{U1_U2_definitions}) must be positive semidefinite.
To see this, we notice that $U_1$ acquires its maxima at $x\rightarrow \pm\infty$ thus is a binding potential.
Obviously, $d\vartheta/dX\equiv 2\sin(\vartheta/2)\sqrt{\cos^2(\vartheta/2)+h}$ is the eigenfunction of $\mathbb{K}_{11}$ with the
eigenvalue $0+k_y^2+k_z^2$ with $k_y$ ($k_z$) coming from a Fourier transform 
in $y$ ($z$) direction.  
On the other hand, $d\vartheta/dX>0$ always holds implying that it has no nodes at any finite $X$.
Therefore $d\vartheta/dX$ must be the ground-state wave function with the lowest
eigenvalue 0 ($k_y=k_z=0$).

Then $\mathbb{K}_{22}$ has to be positive definite to ensure the dynamical stability of 360DWs. 
We rewrite $U_2$ as $U_2=U_1+g(\vartheta)$ with
\begin{equation}\label{g_theta_definition}
g(\vartheta)\equiv\left(\frac{k_{\mathrm{H}}}{k_{\mathrm{E}}}-4h\sin^2\frac{\vartheta}{2}\right)+2d_{\mathrm{i}}\sin\frac{\vartheta}{2}\sqrt{\cos^2\frac{\vartheta}{2}+h}.
\end{equation}
To ensure  $\mathbb{K}_{22}$ to be positive definite, which means that a finite gap
exist between its ground state and zero-energy level, one sufficient but not necessary condition is
that $g(\vartheta)>0$ always hold for any $X$.
This leads to 
\begin{equation}\label{hc_definition}
	h<h_{\mathrm{up}}\equiv\frac{k_{\mathrm{H}}}{k_{\mathrm{E}}}+\frac{d_{\mathrm{i}}^2+d_{\mathrm{i}}\sqrt{d_{\mathrm{i}}^2+4\frac{k_{\mathrm{H}}}{k_{\mathrm{E}}}}}{8}.
\end{equation}
Note that this region is a subset of the real dynamical stability region, but
is the best one that we can now analytically obtain.
In Fig. \ref{fig2}, the $h_{\mathrm{up}}$ with $k_{\mathrm{H}}/k_{\mathrm{E}}=$ 0.0, 0.2, 0.5 and 1.0
are plotted by solid red, green, blue and magenta curves, respectively. 
Interestingly, the static ``absolutely stable region" is the subset of 
all the dynamical stability regions.
This means that the 360DW profile in Eq. (\ref{Profile_of_360DW}) can be
stable under small-amplitude perturbations (for example, spin-wave stimulations)
even when it becomes metastable under $h_c<h<h_{\mathrm{up}}$.

\section{V. Spin wave spectrum under 360DW potentials} 
Within the dynamical stability region, we begin to explore the issue of spin wave spectrum.
By setting $\alpha=0$ and expanding the LLG equation on the basis of $\mathbf{m}_0$
in the power of $\xi$ (small deviation from $\mathbf{m}_0$), we have
\begin{equation}\label{Spin_wave_equation}
	\partial_{\tau} \xi=\Omega\xi,  \qquad \tau\equiv \frac{t}{\mu_0 M_s/(2A\gamma)},
\end{equation}
where
\begin{equation}\label{Omega_operator_definition}
\Omega=\mathbb{J}\mathbb{K}, \qquad \mathbb{J}=\begin{bmatrix}
	0 & -\mathbb{I} \\
	\mathbb{I}  & 0
\end{bmatrix},
\end{equation}
with $\mathbb{I}$ being the identity operator.
Generally, the standard way of solving Eq. (\ref{Spin_wave_equation})  is to find a complete
set of eigenstates of $\Omega$, which are solutions of the spectral equation $\Omega\xi=\nu\xi$.
If they form a complete set, then the general solution is a linear superposition of these eigenmodes.
However, since $\Omega$ is neither Hermitian nor anti-Hermitian, it is not guaranteed 
that a complete set of eigenstates exists, and therefore the general solution of the 
spin wave equation MAYNOT be a linear superposition of these eigenmodes\cite{Laliena_AEM_2022}.

A solvable case is that $\mathbb{K}$ has a strictly positive spectrum.
Furthermore, if $\mathbb{K}_{12}=0$ which is the case in this work,
the situation becomes simpler.
We first define $\Omega_1=\mathbb{K}_{11}$ and $\Omega_2=\mathbb{K}_{22}$.
Then the eigenvalues of $\Omega$ in Eq. (\ref{Spin_wave_equation}) for a (meta)
stable state are complex conjugate pairs of purely imaginary numbers 
(in the simplest case, just one pair). 
Only the zero modes, if exist, can be unpaired.
The $\mathbb{K}$ operator of a (meta)stable state may be gapless or even have zero modes. 
When $\mathbb{K}_{12}=0$, generally the zero and/or gapless modes are
associated to one operator, say $\mathbb{K}_{11}$, and $\mathbb{K}_{22}$ has a finite gap.
Consequently, the appearance of a zero mode in the spectrum of $\mathbb{K}_{22}$
signals the boundary of dynamical stability region, as shown by the $h_{\mathrm{up}}$ curves in Fig. \ref{fig2}.
Within this stability region (excluding the boundary), $\Omega_2$ (also $\Omega_2^{1/2}$) 
must be a Hermitian, positive definite, and invertible operator.
By defining $S=\Omega_2^{1/2}$,  a Hermitian positive semidefinite operator
$\Lambda=\Omega_2^{1/2}\Omega_1\Omega_2^{1/2}$ is related to the non-Hermitian operator $\Omega_2\Omega_1$
by a similarity transformation $\Lambda=S^{-1}\Omega_2\Omega_1 S$, 
hence they have the same spectrum and related eigenfunctions.

Suppose $\psi_j$ ($j$ being a set of labeling index) is one eigenfunction of $\Omega_2\Omega_1$
with eigenvalue $\omega_j^2$ and satisfying $(\psi_j,\Omega_2^{-1}\psi_k)=N_j\delta_{jk}$
where $N_j$ is a normalization constant.
Then the 2D subspace spanned by $\phi_1=(-\psi_j,0)^{\mathrm{T}}$
and  $\phi_2=(0,\Omega_2^{-1}\psi_j)^{\mathrm{T}}$
is invariant under the action of $\Omega$.
Hence $\Omega$ can be diagonalized within it, and the corresponding eigenstates are 
$\xi^{j\sigma}=\phi_1+\mathrm{i}\sigma\omega_j\phi_2=(-\psi_j,\mathrm{i}\sigma\omega_j\Omega_2^{-1}\psi_j)^{\mathrm{T}}$ with the eigenvalue $\mathrm{i}\sigma\omega_j$ and $\sigma=\pm 1$.
The completeness of the eigenstates of $\Lambda$ implies the completeness of the set 
$\psi_j$, and furthermore the set of $\xi^{j\sigma}$. 
Then an arbitrary spin wave vector $\xi=(\xi_1,\xi_2)^{\mathrm{T}}$
can be decomposed as $\xi=\sum_{j\sigma}c_{j\sigma}\xi^{j\sigma}$ with 
$c_{j\sigma}=(-2N_j)^{-1}[(\psi_j,\Omega_2^{-1}\xi_1)+\mathrm{i}\sigma(\psi_j,\xi_2)/\omega_j]$.
Finally, the general solution of the time-dependent wave equation reads
\begin{equation}\label{General_expansion}
	\xi(\mathbf{r},t)=\sum _{j,\sigma}c_{j\sigma}e^{-\mathrm{i}\sigma\omega_j t}\xi^{j\sigma}(\mathbf{r}).
\end{equation}

Based on the above general theory, we move to the issue of spin-wave scattering 
at the 360DW in MHs.
To simplify notation, in this section we replace the reduced dimensionless real coordinates ``$X,Y,Z$"
by the lowercase ``$x,y,z$".
Suppose the cross-section of the MH is large enough compared with the exchange length.
Since the profile of 360DW is only $x-$dependent, we can reasonably assume that 
the $y-$ and $z-$ components of the eigenfunctions of $\Omega_2\Omega_1$ are 
quasi-plane waves with the respective real wave number $k_y$ and $k_z$,
while the $x-$component is an $x-$dependent function which also depends on $k_y$ and $k_z$,
that is
\begin{equation}\label{Eigenfunction_general_form}
	\psi_{k_y,k_z}(x,y,z)=e^{\mathrm{i}k_y y+\mathrm{i} k_z z} \phi_{k_y,k_z} (x).
\end{equation}
The corresponding spectrum equation reads
\begin{equation}\label{Spectrum_equation_real_space}
	\Omega_2\Omega_1\psi_{k_y,k_z}(x,y,z)=\omega^2\psi_{k_y,k_z}(x,y,z),
\end{equation}
which turns to
\begin{equation}\label{Spectrum_equation_k_space}
	\Omega_{2,k_y,k_z}\Omega_{1,k_y,k_z}\phi_{k_y,k_z} (x)=\omega^2\phi_{k_y,k_z} (x),
\end{equation}
where $\omega^2$ now becomes a function of $k_y$ and $k_z$ and 
\begin{equation}\label{Omega12_in_k_space}
	\Omega_{j,k_y,k_z}=k_y^2+k_z^2-\partial_x^2+U_j,\quad j=1,2.
\end{equation}

When $x\rightarrow\pm\infty$, $\cos\vartheta\rightarrow 1$ hence $U_1\rightarrow 1+h$,
$U_2\rightarrow1+h+k_{\mathrm{H}}/k_{\mathrm{E}}$.
The spectrum equation (\ref{Spectrum_equation_k_space}) becomes
\begin{equation}\label{Spectrum_equation_k_space_asymptotic}
	\begin{split}
		&\left(k_y^2+k_z^2-\partial_x^2+1+h+\frac{k_{\mathrm{H}}}{k_{\mathrm{E}}}\right)\left(k_y^2+k_z^2-\partial_x^2+1+h\right)\phi_{k_y,k_z} \\
		&=\omega^2\phi_{k_y,k_z}.
	\end{split}
\end{equation}
Generally its solution is an exponential function that is written as $e^{\mathrm{i}k_z z}$,
where $k_z$ can be either real or complex. 
Hence one has $\omega^2=\omega_1\omega_2$ with
$\omega_1=k^2+1+h$, $\omega_2=\omega_1+k_{\mathrm{H}}/k_{\mathrm{E}}$ 
and $k^2\equiv k_x^2+k_y^2+k_z^2$.
We can solve the $k_x(\omega)$ relation by inverting the above result as
\begin{equation}\label{k_x_omega_relation_x_infty}
	k_x^2=-\left(1+h+\frac{k_{\mathrm{H}}}{2k_{\mathrm{E}}}+k_y^2+k_z^2\right)\pm\sqrt{\left(\frac{k_{\mathrm{H}}}{2k_{\mathrm{E}}}\right)^2+\omega^2},
\end{equation}
which means that $k_x^2$ must be real hence $k_x$ can only be either real or pure imaginary.

When $k_x$ is pure imaginary ($k_x^2<0$), the state is a bound state in $x-$direction.
There are two possibilities:
(i) the minus sign is taken in Eq. (\ref{k_x_omega_relation_x_infty})
thus no restriction is performed on $\omega$.
(ii) the plus sign is taken meantime maintains $k_x^2<0$ thus $\omega$ has an upper limit, that is,
$\omega<\omega_{\mathrm{G}}(k_y,k_z)$ with
\begin{equation}\label{omega_G_definition}
	\omega_{\mathrm{G}}=\sqrt{\left(1+h+\frac{k_{\mathrm{H}}}{k_{\mathrm{E}}}+k_y^2+k_z^2\right)\left(1+h+k_y^2+k_z^2\right)}.
\end{equation}
Consequently, the only possibility for a state being a continuum state is that 
the plus sign is taken in Eq. (\ref{k_x_omega_relation_x_infty}) and 
$\omega>\omega_{\mathrm{G}}(k_y,k_z)$ so as to ensure a real $k_x$.
Therefore, continuum states have a finite gap above the zero frequency.
Meantime,  Eq. (\ref{k_x_omega_relation_x_infty}) with the plus sign provides
the dispersion relation for the continuum states, which will be our main concern
in exploring the SWGHE in MHs since the incident, reflected and transmitted spin waves
should all be continuum states coming from and/or going to infinity.

On the other hand, since $\Omega_{2,k_y,k_z}\Omega_{1,k_y,k_z}$ is real and
unchanged under the parity transformation $x\rightarrow -x$, 
its eigenfunctions can be grouped to be real and with definite parity 
(either even or odd functions under  $x\rightarrow -x$).
This also holds for their asymptotic behaviors for $x\rightarrow \pm\infty$.
Back to the continuum states we are focusing on. 
They therefore can be labeled by $k_z(\ge 0)$ and the parity which is denoted by 
the symbols ``$e$ (even)" and ``$o$ (odd)". 
Accordingly, we write $\phi_{k_x,k_y,k_z}^{(e)}$ and $\phi_{k_x,k_y,k_z}^{(o)}$ 
as the continuum eigenfunctions of $\Omega_{2,k_y,k_z}\Omega_{1,k_y,k_z}$ with
even and odd parities, respectively. 
Their asymptotic behaviors for $x\rightarrow\pm\infty$ are
\begin{equation}\label{phi_e_o_asymptotic}
	\begin{split}
    	\phi_{k_x,k_y,k_z}^{(e)}(x\rightarrow\pm\infty)&\sim\cos\left(k_x x\pm\delta_e\right), \\
    	\phi_{k_x,k_y,k_z}^{(o)}(x\rightarrow\pm\infty)&\sim\sin\left(k_x x\pm\delta_o\right),
	\end{split}
\end{equation}
where the nonzero phase shifts $\delta_{e,o}$ depends on $k_{x,y,z}$ in general.
Also note that opposite signs in front of $\delta_{e,o}$ are taken when $x$ approaches 
the left and right end of the MH so as to fulfill the symmetric demands imposed on 
$\phi_{k_x,k_y,k_z}^{(e)}$ and $\phi_{k_x,k_y,k_z}^{(o)}$, respectively.

Based on the above results, we construct the eigenfunctions of 
$\Omega$ for $x\rightarrow\pm\infty$ as follows.
Suppose the index $j$ denote the combination $(k_x,k_y,k_z,p)$ with $p=e$ or $o$, 
thus the eigenstate of $\Omega_{2,k_y,k_z}\Omega_{1,k_y,k_z}$ is 
$\psi_j=e^{\mathrm{i}k_y y+\mathrm{i} k_z z} \phi_{k_x,k_y,k_z}^{(p)}$.
At $x\rightarrow\pm\infty$, one has 
$\Omega_{2,k_y,k_z}=k_y^2+k_z^2-\partial_x^2+1+h+k_{\mathrm{H}}/k_{\mathrm{E}}$, thus
$\Omega_{2,k_y,k_z}^{-1}\psi_j=(k^2+1+h+k_{\mathrm{H}}/k_{\mathrm{E}})^{-1}e^{\mathrm{i}k_y y+\mathrm{i} k_z z} \phi_{k_x,k_y,k_z}^{(p)}=e^{\mathrm{i}k_y y+\mathrm{i} k_z z} \phi_{k_x,k_y,k_z}^{(p)}/\omega_2$.
On the other hand, $\omega_{k_x,k_y,k_z,p}=\sqrt{\omega_1\omega_2}$.
Therefore, we have the following explicit form of the eigenstate of the $2\times2$ operator $\Omega$
under the potentials of our 360DW at $x\rightarrow\pm\infty$ as
\begin{equation}\label{Eigenstate_of_Omega_360DW}
	\xi_{k_x,k_y,k_z}^{(p,\sigma)}=e^{\mathrm{i}k_y y+\mathrm{i} k_z z}\phi_{k_x,k_y,k_z}^{(p)}\Xi^{(\sigma)}, \quad
	\Xi^{(\sigma)}\equiv\begin{bmatrix}
		-1 \\
		\mathrm{i}\sigma\sqrt{\frac{\omega_1}{\omega_2}}
	\end{bmatrix}.
\end{equation}
Finally, the general asymptotic solution of the wave equation involving only continuum states 
turns to be
\begin{equation}\label{Spin_wave_general_expansion}
	\xi(\mathbf{r},t)=\sum_{p\sigma}\iint\frac{dk_ydk_z}{(2\pi)^2}\int_{0}^{+\infty}dk_x c_{p\sigma}e^{\mathrm{i}(k_y y+k_z z-\sigma\omega\tau)}\xi_{k_x,k_y,k_z}^{(p,\sigma)},
\end{equation}
where $\sigma=\pm1$, $p=e,o$ and  $c_{p\sigma}$ is the 
expanding coefficient and generally is a function of the wave numbers $k_{x,y,z}$.
Note that there is no need to introduce the Hermitian conjugate part (c.c.) here, 
because the spin wave components can be complex in general.
At last, remember that here our real space is the reduced dimensionless one. 
When projecting our theory to real MHs, a ``$\Delta_0$" factor should be appended to ``$x,y,z$"
(indeed ``$X,Y,Z$") so as to restore their length dimension.

\section{VI. Spin wave scattering by 360DWs - the SWGHE} 
With the asymptotic properties of the wave operator and the completeness of its eigenstates
presented in the above section, we hereby explore the scattering behaviors of spin waves
at 360DWs, that is, the SWGHE, by the standard scattering theory.

As shown in Fig. \ref{fig1}, a spin-wave package with sharp peak 
around the wave number $k_{x0}>0,k_{y0}>0,k_{z0}>0$ is originally located at
$(x,y,z)=(-\infty,-\infty,-\infty)$ when $\tau\rightarrow-\infty$.
It moves towards the 360DW at $x=0$ and scattered by it at $\tau=0$.
As products of this scattering, when $\tau\rightarrow +\infty$ two outgoing wave packages
emerge: a reflected one moves towards the $(x,y,z)=(-\infty,+\infty,+\infty)$ region, 
and a transmitted one moves towards the $(x,y,z)=(+\infty,+\infty,+\infty)$ region. 
They also have the sharp peak around the wave number $(k_{x0},k_{y0},k_{z0})$.
In principle these three wave packages can be exactly described by 
Eq. (\ref{Spin_wave_general_expansion}) with suitable universal coefficients 
$c_{p\sigma}$ across different asymptotic regions.
On the other hand, physically in the $x\rightarrow -\infty$ region the incident
and reflected wave packages coexist thus one has
\begin{widetext} 
\begin{equation}\label{Spin_wave_incident_and_reflected_waves}
	\xi_{x\rightarrow-\infty}(\mathbf{r},t)=\iint\frac{dk_ydk_z}{(2\pi)^2}\int_{0}^{+\infty}dk_x\left[A e^{\mathrm{i}(k_y y+k_z z+k_x x-\omega\tau)}+A_{R} e^{\mathrm{i}(k_y y+k_z z-k_x x-\omega\tau)}\right] \Xi^{(+)},
\end{equation}
\end{widetext} 
in which $A$ ($A_R$) is the amplitude of incident (reflected) wave and
is a function of $(k_x,k_y,k_z)$.
Meantime only $\sigma=+1$ terms, that is $\Xi^{(+)}$, are involved since the phase velocities 
in $y,z$ directions should be positive.
Alternatively, in the $x\rightarrow +\infty$ region only transmitted wave exists thus
\begin{equation}\label{Spin_wave_transmitted_wave}
	\xi_{x\rightarrow+\infty}(\mathbf{r},t)=\iint\frac{dk_ydk_z}{(2\pi)^2}\int_{0}^{+\infty}dk_x A_{T} e^{\mathrm{i}(k_y y+k_z z+k_x x-\omega\tau)}\Xi^{(+)},
\end{equation}
with $A_T$ being the wave-number-dependent amplitude of the transmitted wave.
By comparing Eqs. (\ref{Spin_wave_incident_and_reflected_waves}) and
(\ref{Spin_wave_transmitted_wave}) with the general expansion 
in Eq. (\ref{Spin_wave_general_expansion}) as well as the asymptotic behaviors 
in Eq. (\ref{phi_e_o_asymptotic}), 
we have 
\begin{equation}\label{AR_AT_depend_on_A}
\begin{split}
	A_R&=\left[e^{\mathrm{i}\left(\delta_e+\delta_o+\frac{\pi}{2}\right)}\sin\left(\delta_e-\delta_o\right)\right]A\equiv\mathcal{R}A,   \\
	A_T&=\left[e^{\mathrm{i}\left(\delta_e+\delta_o\right)}\cos\left(\delta_e-\delta_o\right)\right]A\equiv\mathcal{T}A.
\end{split}
\end{equation}
The identity $|\mathcal{R}|^2+|\mathcal{T}|^2\equiv1$ describes the conservation 
of magnetic energy when neglecting the Gilbert damping in this magnetic system
since we have set $\alpha=0$ in deducing Eq. (\ref{Spin_wave_equation}).
In addition, The extra $\pi/2$ phase in $A_R$ implies the different 
polarization behaviors between the reflected and transmitted spin waves.

We now turn to the scattering behaviors of spin waves by the 360DW.
It is well understood that when $\tau\rightarrow\pm\infty$,
generally the phases oscillate very rapidly and destructive interference dominates in most cases.
On the contrary, constructive interference takes place (thus significant wave intensities emerge) 
only at the points where the phase is stationary 
with respect to $k_{x,y,z}$, which is the central spirit for exploring the SWGHE.
Specifically, we first write $A=|A|e^{\mathrm{i}\phi_A}$.
When $\tau\rightarrow-\infty$, from Eq. (\ref{Spin_wave_incident_and_reflected_waves})
the total phase of the incident spin wave is 
$\Phi_{\mathrm{in}}=\phi_A+\sum_{j}k_{x_j} x_j-\omega \tau$
with $x_1\equiv x$, $x_2\equiv y$ and $x_3\equiv z$.
By demanding that this phase is constant, the trace line of the observable incident spin wave is
$x_j=-\partial\phi_A/\partial k_{x_j}+(\partial\omega/\partial k_{x_j})\tau$.
It passes through the point $P_{\mathrm{in}}=(-\partial\phi_A/\partial k_x,-\partial\phi_A/\partial k_y,-\partial\phi_A/\partial k_z)$ 
and is parallel with the orientation vector 
$\mathbf{V}\equiv(\partial\omega/\partial k_x,\partial\omega/\partial k_y,\partial\omega/\partial k_z)$ in real space, 
thus can be rewritten as
\begin{equation}\label{Trace_line_incident_wave}
	\frac{x+\partial\phi_A/\partial k_x}{\partial\omega/\partial k_x}=\frac{y+\partial\phi_A/\partial k_y}{\partial\omega/\partial k_y}=\frac{z+\partial\phi_A/\partial k_z}{\partial\omega/\partial k_z}.
\end{equation}
Obviously, the intersection of this trace line with the $x=0$ plane (where 360DW locates) is
\begin{equation}\label{Intersection_of_incident_wave_with_xeq0}
	y_{\mathrm{in}}=\frac{\partial\omega/\partial k_y}{\partial\omega/\partial k_x}\frac{\partial\phi_A}{\partial k_x}-\frac{\partial\phi_A}{\partial k_y},\ \  z_{\mathrm{in}}=\frac{\partial\omega/\partial k_z}{\partial\omega/\partial k_x}\frac{\partial\phi_A}{\partial k_x}-\frac{\partial\phi_A}{\partial k_z}.
\end{equation}
In particular, if $\phi_A$ is constant (which is common in real experiments), 
the trace line goes through the origin of coordinate system.

On the other hand, when $\tau\rightarrow+\infty$, the total phase of the transmitted spin wave 
turns to be $\Phi_{\mathrm{tr}}=\phi_A+(\delta_e+\delta_o)+\sum_{j}k_{x_j} x_j-\omega \tau$,
as shown by Eqs. (\ref{Spin_wave_transmitted_wave}) and (\ref{AR_AT_depend_on_A}).
Again the ``constant phase" condition requests that the trace line of the observable 
transmitted wave is
$x_j=-\partial\phi_A/\partial k_{x_j}-\partial(\delta_e+\delta_o)/\partial k_{x_j}+(\partial\omega/\partial k_{x_j})\tau$, that is
\begin{equation}\label{Trace_line_transmitted_wave}
	\begin{split}
		&\frac{x+\partial\phi_A/\partial k_x+\partial(\delta_e+\delta_o)/\partial k_x}{\partial\omega/\partial k_x}  \\
		=&\frac{y+\partial\phi_A/\partial k_y+\partial(\delta_e+\delta_o)/\partial k_y}{\partial\omega/\partial k_y} \\
		=&\frac{z+\partial\phi_A/\partial k_z+\partial(\delta_e+\delta_o)/\partial k_z}{\partial\omega/\partial k_z},
	\end{split}
\end{equation}
which is parallel to that of the incident wave but in principle not coincident with it. 
If the width of 360DW is much narrower than the wave length of spin waves, 
the intersection of this trace line with the $x=0$ plane, that is
\begin{equation}\label{Intersection_of_transmitted_wave_with_xeq0}
	\begin{split}
		y_{\mathrm{tr}}&=\frac{\partial\omega/\partial k_y}{\partial\omega/\partial k_x}\left[\frac{\partial\phi_A}{\partial k_x}+\frac{\partial(\delta_e+\delta_o)}{\partial k_x}\right]-\frac{\partial\phi_A}{\partial k_y}-\frac{\partial(\delta_e+\delta_o)}{\partial k_y},  \\  z_{\mathrm{tr}}&=\frac{\partial\omega/\partial k_z}{\partial\omega/\partial k_x}\left[\frac{\partial\phi_A}{\partial k_x}+\frac{\partial(\delta_e+\delta_o)}{\partial k_x}\right]-\frac{\partial\phi_A}{\partial k_z}-\frac{\partial(\delta_e+\delta_o)}{\partial k_z},
	\end{split}
\end{equation}
can be viewed as the exit point of the transmitted wave.
Hence, a lateral displacement in the center of transmitted wave packet with respect to 
that of the incident wave (if there was no 360DW) emerges and turns to be
\begin{equation}\label{GH_lateral_shift_transmitted_wave_general}
	\begin{split}
		\Delta y_{\mathrm{GH}}&=y_{\mathrm{tr}}-y_{\mathrm{in}}=\frac{\partial\omega/\partial k_y}{\partial\omega/\partial k_x}\frac{\partial(\delta_e+\delta_o)}{\partial k_x}-\frac{\partial(\delta_e+\delta_o)}{\partial k_y},\\
		\Delta z_{\mathrm{GH}}&=z_{\mathrm{tr}}-z_{\mathrm{in}}=\frac{\partial\omega/\partial k_z}{\partial\omega/\partial k_x}\frac{\partial(\delta_e+\delta_o)}{\partial k_x}-\frac{\partial(\delta_e+\delta_o)}{\partial k_z}.
	\end{split}
\end{equation}
The dispersion relationship in Eq. (\ref{Spectrum_equation_k_space_asymptotic})
provides $\partial \omega/\partial k_j=k_j(2\omega_1+ k_{\mathrm{H}}/k_{\mathrm{E}})/\omega$
with $j\in\{x,y,z\}$.
Therefore, the above lateral displacements turn to
\begin{equation}\label{GH_lateral_shift_transmitted_wave_final}
	\begin{split}
		\Delta y_{\mathrm{GH}}&=\frac{k_y}{k_x}\frac{\partial(\delta_e+\delta_o)}{\partial k_x}-\frac{\partial(\delta_e+\delta_o)}{\partial k_y},\\
		\Delta z_{\mathrm{GH}}&=\frac{k_z}{k_x}\frac{\partial(\delta_e+\delta_o)}{\partial k_x}-\frac{\partial(\delta_e+\delta_o)}{\partial k_z}.
	\end{split}
\end{equation}
Two interesting inferences can be drawn from the above equation.
(i) Only when the spin wave is vertically incident ($k_y=k_z=0$), the Goos-H\"{a}nchen lateral displacements
restores to the existing result in Ref. \cite{Laliena_AEM_2022} as
\begin{equation}\label{GH_lateral_shift_transmitted_wave_vertically_incident}
	\begin{split}
		\Delta y_{\mathrm{GH}}=-\frac{\partial(\delta_e+\delta_o)}{\partial k_y},\quad
		\Delta z_{\mathrm{GH}}=-\frac{\partial(\delta_e+\delta_o)}{\partial k_z}.
	\end{split}
\end{equation}
(ii) Generally the spin wave is obliquely incident with an incident angle $\alpha>0$. 
After defining $k_y/k_x\equiv\tan\alpha_1$ and $k_z/k_x\equiv\tan\alpha_2$
 ($0\le\alpha_{1,2}<\pi/2$ and $\tan^2\alpha_1+\tan^2\alpha_2=\tan^2\alpha$), 
 one has $\partial/\partial k_y=(\cot\alpha_1)\partial/\partial k_x$ and
 $\partial/\partial k_z=(\cot\alpha_2)\partial/\partial k_x$.
 Then the Goos-H\"{a}nchen lateral displacements finally become
\begin{equation}\label{GH_lateral_shift_transmitted_wave_obliquely_incident}
	\begin{split}
		\Delta y_{\mathrm{GH}}&=\left(\tan\alpha_1-\cot\alpha_1\right)\frac{\partial(\delta_e+\delta_o)}{\partial k_x},\\
		\Delta z_{\mathrm{GH}}&=\left(\tan\alpha_2-\cot\alpha_2\right)\frac{\partial(\delta_e+\delta_o)}{\partial k_x}.
	\end{split}
\end{equation} 
Interestingly, when $\alpha_1=\pi/4$ ($\alpha_2=\pi/4$) the lateral displacement 
$\Delta y_{\mathrm{GH}}$ ($\Delta z_{\mathrm{GH}}$) vanishes.

At last, for reflected wave, from Eqs. (\ref{Spin_wave_incident_and_reflected_waves}) and
(\ref{AR_AT_depend_on_A}) the corresponding total phase reads
$\Phi_{\mathrm{rfl}}=\phi_A+(\delta_e+\delta_o+\pi/2)+k_y y+k_z z-k_x x-\omega \tau$.
The ``constant phase" request results in the trace line of observable reflected wave as
\begin{equation}\label{Trace_line_reflected_wave}
	\begin{split}
		&\frac{x-\partial\phi_A/\partial k_x-\partial(\delta_e+\delta_o)/\partial k_x}{\partial\omega/\partial k_x}  \\
		=&\frac{y+\partial\phi_A/\partial k_y+\partial(\delta_e+\delta_o)/\partial k_y}{\partial\omega/\partial k_y} \\
		=&\frac{z+\partial\phi_A/\partial k_z+\partial(\delta_e+\delta_o)/\partial k_z}{\partial\omega/\partial k_z}.
	\end{split}
\end{equation}
This trace line, together with the transmitted one shown in Eq. (\ref{Trace_line_transmitted_wave}) are 
mirror-symmetric with respect to the $x=0$ plane.
Therefore, the intersection of this trace line with the $x=0$ plane is the same as
that between the transmitted wave and the $x=0$ plane, hence leading to the same
lateral displacements as those in Eq. (\ref{GH_lateral_shift_transmitted_wave_general}) and further in Eq. 
(\ref{GH_lateral_shift_transmitted_wave_final}).

\begin{figure} [htbp]
	\centering
	\includegraphics[width=0.44\textwidth]{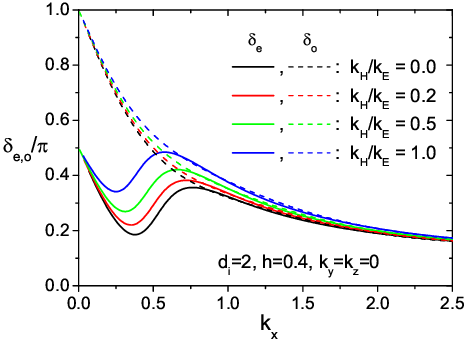}
	\caption{(Color online) Phase shifts $\delta_{e,o}$ of the eigenfunctions $\phi_{k_x,k_y,k_z}^{(e,o)}(|x|\rightarrow\infty)$ 
		under $d_{\mathrm{i}}=2$, $h=0.4$, $k_y=k_z=0$ and 
		various $k_{\mathrm{H}}/k_{\mathrm{E}}$ ($=0.0$, 0.2, 0.5 and 1.0) which all fall into
		the dynamical stability region in Fig. \ref{fig2}.
		The solid and dashed curves are for the phase shifts with even and odd parities, respectively.}\label{fig3}
\end{figure}

At the end of this section, we perform some numerical calculations on the phase shifts $\delta_{e,o}$
and the resulting Goos-H\"{a}nchen lateral displacements.
The brief calculation procedure is as follows.
We first numerically solve the spectrum equation [Eq. (\ref{Spectrum_equation_k_space})]
for a fixed combination $\{k_y,k_z\}$ belonging to a dense-enough discrete 2D wave-number set
on a large box $-L\le x \le L$ with the boundary condition $\phi_{k_y,k_z}(x=\pm L)=0$, 
and then obtain the eigenvalues $\omega^2(k_x,k_y,k_z)$ and the corresponding 
eigenfunctions $\phi_{k_x,k_y,k_z}^{(e,o)}(x)$ with definite parity ``$(e,o)$".
Note that the real wave number $k_x$ in $\omega^2(k_x,k_y,k_z)$ and $\phi_{k_x,k_y,k_z}^{(e,o)}(x)$ is obtained 
by Eq. (\ref{k_x_omega_relation_x_infty}) with ``plus" sign for this certain combination $\{k_y,k_z\}$.
The boundary condition $\phi_{k_x,k_y,k_z}^{(e,o)}(x=\pm L)=0$ as well as the asymptotic forms 
in Eq. (\ref{phi_e_o_asymptotic}) demand that $\cos(k_x L+\delta_e)=0$ and $\sin(k_x L+\delta_o)=0$.
This leads to $k_x L+\delta_e=(n_e+1/2)\pi$ and $k_x L+\delta_o=n_o\pi$ where $n_{e,o}$
are integers that ensure $0\le \delta_{e,o}\le \pi$.
Therefore, one has $\delta_e=(-k_x L)\ \mathrm{mod}\ \pi+\pi/2$ and $\delta_o=(-k_x L)\ \mathrm{mod}\ \pi$.
Note that there is no ``$\delta_e=\delta_o\pm\pi/2$" identity for every $k_x$ 
because generally the wave number $k_x$ for $\phi_{k_y,k_z}^{(e)}(x)$ and $\phi_{k_y,k_z}^{(o)}(x)$ are different.
Finally, the above procedure has been repeated for quite a few box lengths and differential step sizes 
to exclude noticeable volume or discretization effects.

In Fig. \ref{fig3}, the phase shifts $\delta_{e}$ and $\delta_{o}$ of the eigenfunctions 
$\phi_{k_x,k_y,k_z}^{(e,o)}(x)$ at $x\rightarrow \pm\infty$ are provided by solid and dashed
curves, respectively.
In the calculations, $d_{\mathrm{i}}=2$, $h=0.4$ and $k_y=k_z=0$. 
Four reasonable values of $k_{\mathrm{H}}/k_{\mathrm{E}}$, 
which are 0.0, 0.2, 0.5 and 1.0, are selected and respectively marked by black, red, green and blue symbols.
All these phase points fall into the corresponding dynamical stability region in Fig. \ref{fig2}.
For small wave numbers (long-wavelength spin waves), $\delta_o$ is larger than $\delta_e$ 
by an approximate increment of $\pi/2$.
However, as $k_x$ increases to 2.5, $\delta_{e,o}$ tend to converge to the same value, even
under different $k_{\mathrm{H}}/k_{\mathrm{E}}$.
In addition, $\delta_o$ monotonically decreases from $\pi$ to 0, while $\delta_e$ first suffers
a non-monotonic twist and eventually converges to $\delta_o$.
This is distinct from the counterparts in the case of SWGHE at a Sine-Gordon soliton
in an ideal 1D monoaxial helimagnet, which should be the consequence of different
potentials induced by the Sine-Gordon soliton and our 360DW herein.

\begin{figure} [htbp]
	\centering
	\includegraphics[width=0.44\textwidth]{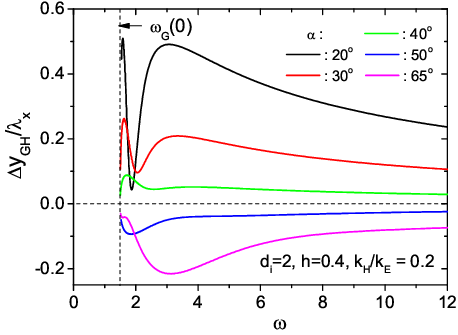}
	\caption{(Color online) Goos-H\"{a}nchen lateral displacements of spin wave beams
		lying in $xy-$plane and with several incident angles (in degree).
		The phase point is $d_{\mathrm{i}}=2$, $h=0.4$, and $k_{\mathrm{H}}/k_{\mathrm{E}}=0.2$ 
		that falls into	the dynamical stability region in Fig. \ref{fig2}.
		The vertical dash line indicates $\omega_{\mathrm{G}}(0)$.}\label{fig4}
\end{figure}

With the phase shifts in hand and by taking the corresponding partial differentials with respect
to $k_{x,y,z}$, the Goos-H\"{a}nchen lateral displacements can be obtained.
In numerical calculations, we consider the case where incident spin waves lie in the $xy-$plane 
with the incident angle $\alpha$.
Hence $k_z\equiv 0$ and from Eq. (\ref{GH_lateral_shift_transmitted_wave_obliquely_incident})
one has finite $\Delta y_{\mathrm{GH}}$ and disappearing $\Delta z_{\mathrm{GH}}$.
In Fig. \ref{fig4}, the eigenfrequency dependence of $\Delta y_{\mathrm{GH}}$ (in the
unit of wavelength $\lambda_x$ in $x-$direction) for different incident angle is provided.
In the calculations, $d_{\mathrm{i}}=2$, $h=0.4$, and $k_{\mathrm{H}}/k_{\mathrm{E}}=0.2$,
making the corresponding phase point lie within the dynamical stability region of 360DWs.
The vertical dash line indicates $\omega_{\mathrm{G}}(0)$ which is the lower limit of 
continuum-state frequency with $k_y=k_z=0$.
Numerical data provide that when $\alpha$ increases from $20^{\circ}$ to $65^{\circ}$,
Goos-H\"{a}nchen lateral displacement decreases and even reverse its direction 
when $\alpha$ exceeds $45^{\circ}$.
All results show that $\Delta y_{\mathrm{GH}}$ is a fraction of $\lambda_x$
and thus can be enhanced to several or even tens of wavelength when
the spin wave passes through a CSL.
On the other hand, as $\omega$ increases from $\omega_{\mathrm{G}}(0)$ the lateral displacement 
suffers a more complicated non-monotonous behavior compared with that in the case of an
isolated Sine-Gordon soliton in a monoaxial helimagnet.
This can be attributed to the nontrivial monotonous feature of the $U_{2}$ potential induced by the 360DW profile 
in Eq. (\ref{Profile_of_360DW}).

\section{VII. Discussions and conclusion} 
First of all, during our analytical exploration we have neglected the nonlocal component 
of the magnetostatic energy. 
This results in the ``decorated" magnetic anisotropy coefficients
$k_{\mathrm{E}}=k_1+(D_x-D_z)$ and $k_{\mathrm{H}}=k_2+(D_y-D_x)$ 
with $k_1$ ($k_2$) being the crystalline anisotropy coefficient in the easy 
$z-$ (hard $y-$) axis and $D_{x,y,z}$ being the three average demagnetization factors.
Consequently, we obtain the rigorous profile of 360DWs in Eq. (\ref{Profile_of_360DW})
which leads to the unique potentials in Eq. (\ref{U1_U2_definitions}).
These potentials are fundamental for the noncommutativity of $\Omega_1$ and $\Omega_2$
and furthermore the finite Goos-H\"{a}nchen lateral displacements.
When the nonlocal component of the magnetostatic energy is considered, 
the anisotropy energy then can no longer be simply expressed as quadratic terms
of the magnetization components, hence the analytical profile 
of 360DWs [for example, Eq. (\ref{Profile_of_360DW})] can hardly be obtained.
However, we can reasonably infer that the operators $\mathbb{K}_{\alpha\beta}$ 
will depend on $k_{y,z}$ in a nontrivial way thus further confirms the emergence of SWGHE.
The left question is whether the inclusion of the nonlocal component enhances 
or weakens the magnitude of SWGHE, which needs further explorations.

Second, we reemphasize the importance of the existence of dynamical stability
region for 360DWs with the profile shown in Eq. (\ref{Profile_of_360DW}), 
which is the central result of this work.
Within it [see Eq. (\ref{hc_definition})], the operator $\mathbb{K}_{22}$ is positive definite 
thus the spin wave
propagation and scattering problem can be solved in the non-perturbation framework
as described in the main text.
Also, it has several important implications. 
(i) Obviously, even in the absence of IDMI ($d_{\mathrm{i}}=0$), the 360DW can 
still be dynamically stable when $h<k_{\mathrm{H}}/k_{\mathrm{E}}$.
This holds even in the case of uniaxial FM layers in real MHs [$k_2=0$ but $k_{\mathrm{H}}>0$]
or infinite biaxial FM materials with easy (hard) axis being along $z$ ($y$) axis.
On the contrary, for a 1D 360DW extending in $x-$axis but homogeneous in
$y-$ and $z-$axes of an infinite FM material with unixial anisotropy in $z-$axis,
this 360DW MAY be dynamically unstable and could be destroyed by a spin wave flowing through it.
(ii) The presence of IDMI enlarges the dynamical stability region of the 360DW.
Even in the case of $k_{\mathrm{H}}=0$, the 1D 360DW will be dynamically stable
under the disturbance of spin waves with small amplitude.
(iii) As expected, the operators $\Omega_1$ and $\Omega_2$ are noncommutative in the
presence of IDMI, thus leads to the finite Goos-H\"{a}nchen lateral displacements
which confirms the assertion in Ref. \cite{Laliena_AEM_2022}. 
However, Eq. (\ref{g_theta_definition}) provides that even in the absence of IDMI
$\Omega_1$ and $\Omega_2$ can still be noncommutative
due to the unique potentials induced by the 360DW in Eq. (\ref{Profile_of_360DW}),
which is distinct to the most-commonly-encountered P\"{o}schl-Teller ones.
This delivers novel possibilities of observing SWGHE in real MHs.

Third, the lateral-displacement results in Eq. (\ref{GH_lateral_shift_transmitted_wave_final}) 
implies several important inferences. 
(i) When the spin wave is vertically incident to the 360DW, $k_y=k_z=0$ hence 
$\Delta y_{\mathrm{GH}}$ and $\Delta z_{\mathrm{GH}}$ reproduce the results in Ref. \cite{Laliena_AEM_2022}. 
The emergence of these nonzero lateral displacements relies on the fact that $\delta_{e,o}$ must
depend on $k_{y,z}$, which has been asserted as the necessary condition for the existence 
of SWGHE even for infinite dimensions in the transverse plane.
However, it should be noted that the partial differentials at this time need to be taken 
around $k_y=k_z=0$.
(ii) Equation (\ref{GH_lateral_shift_transmitted_wave_final}) provides an alternative possibility
for the emergence of SWGHE: an oblique incident spin wave as long as $\delta_{e,o}$ 
depend on $k_{x}$. 
In fact, the discussions for the case of a Sine-Gordon soliton in a monoaxial helimagnet 
in Ref. \cite{Laliena_AEM_2022} should deliver the same conclusion, however the lateral displacements they obtained 
[Eq. (52) therein] are distinct from our results above.

Forth, in real MHs the finite dimensions in $y$ and $z$ directions generally
diminish the feasibility of introducing the wave numbers $k_y$ and $k_z$.
However, for MHs with transverse dimensions quite larger than the wave length
of spin waves, it is an acceptable way of dealing with this issue.
Not to mention that in the absence of IDMI and for large enough biaxial FM materials with easy (hard) 
axis being along $z$ ($y$) axis, a 360DW with the profile described 
by Eq. (\ref{Profile_of_360DW}) and extending in $x-$direction can 
fully fulfill the technique introduced in the main text.

In summary, the SWGHE at 360DWs in MHs has been systematically explored
on the basis of a thorough investigation on the static and dynamical stability of 360DWs.
In addition to the predictable important role of IDMI in dynamically stabilizing 360DWs and
efficiently inducing the lateral displacements of spin waves at 360DWs, 
we highlight the survival of SWGHE in ferromagnets with biaxial anisotropy 
(either intrinsic or caused by shape anisotropy) in 
the absence of IDMI due to the unique 360DW-induced potentials which are distinct 
to the well-known P\"{o}schl-Teller ones.
In addition, the lateral displacements of vertically or obliquely incident spin waves at 360DWs results from
different dependence of phase shifts $\delta_{e,o}$ on the wave numbers $k_{x,y,z}$,
which has not been revealed before.
Our results pave the way for development and performance optimization
in future magnetic nandevices based on spin wave propagation and scattering at
360DWs in real MHs.

\section{acknowledgments} 
M.L. acknowledges support from the National Natural Science Foundation of China (Grant No. 12204403).
B.X. is funded by the National Natural Science Foundation of China (Grant No. 11774300).
J.L. is supported by the Natural Science Foundation of Jiangsu Province.

%
%


\end{document}